# Information in Quantum Description and Gate Implementation

Gayathre Krishnan

**Abstract:** This note considers Kak's observer-reference model of quantum information, where it is shown that qubits carry information that is $\frac{\sqrt{n}}{\ln n}$ times classical information, where $n$ is the number of components of the measurement system, to analyze information processing in quantum gates. The obverse side of this exponential nature of quantum information is that the computational complexity of implementing unconditionally reliable quantum gates is also exponential.

**Introduction**

Although quantum information theory considers ensembles of qubits in about the same manner as classical information, there are unique aspects of the quantum measurement process that indicate that this picture is not entirely satisfactory. This is clear not only in communications, but also in the teleportation protocol [1]. The particular difficulty in the communication of an unknown state is that there is no means to check if what has been received is what had really been transmitted, which defines a scenario unlike what is found in classical communication.

Subhash Kak has argued [2] that the amount of information that can be extracted from a qubit depends on the measurement apparatus. He took a specific arrangement of horizontally and vertically polarized photons that were passed through a distributed measurement apparatus. He concluded that a quantum measurement gives information roughly

$$\frac{\sqrt{n}}{\ln n} \times \text{Classical information} \qquad (1)$$

As $n$ (the number of components in the system) becomes large, the advantage over classical information becomes enormous. This means that for the same performance, one would need of the order of $2^{\sqrt{n}}$ classical elements where only $n$ quantum elements would have the same performance.

But the reverse view of the quantum information question is that since it is being generated in the quantum circuit, in which the control will be provided by classical variables, the implementation of the quantum circuit is going to be exponentially harder than that of a classical circuit.

There are some studies on actual gate implementation, but here we mention only two as representative of the work. Ivanyos, Massar, and Nagy [3] studied the computation power of lattices composed of two dimensional systems (qubits) on which translationally invariant global two-qubit gates can be performed. Their implementation was subject to the realizability of a set of 6 global two qubit gates. They argued that "the global two-qubit gate considered has almost as many eigenvalues as possible. But this means that in some sense this gate acts 'chaotically'. Therefore the model is probably not very useful in the sense that it does not seem to allow one to define a qubit structure on the eigenspace in a natural way."

Xiao and Jones [4] have used a system of composite pulses to suppress errors on one qubit and two qubit gates in an NMR implementation. In our view, the results are disappointing since even their most promising approach called the BB1 has severe limitations. "The more complex B4 and P4 sequences, although theoretically superior, do not perform well in practice."

In this article, we examine implications of equation (1), while noting that the advantage of quantum processing is predicated on an appropriately "located" observer, and discuss the problem of the implementation of quantum gates.

**Information and uncertainty**

The standard quantum circuit model of quantum information processing defines an analog framework. But doing so opens up the model to the problems related to noise and uncertainty [5]. There is, furthermore, the question of the role decoherence plays in the collapse of the quantum state [6]. Given that the processing must be done with unknown states, it is a challenge to define appropriate protocols to test the circuits. As argued in [1], the information transfer cannot be defined unconditionally. It is difficult then to perform many basic quantum gate operations reliably [7], as argued by Ponnath [8].

**Complexity of gates**

If the gate is a physical device then from an information point of view its control can be characterized in terms of entropy.

Computation with noisy components would require that the quantum circuit is associated with an entropy rate smaller than the information capacity of the controller used. But since quantum information is exponential, it is computationally infeasible for a classical controller to match the quantum entropy of the circuit.

Taking the case of a CNOT circuit associated with specific faults, Kak has shown [9] that it becomes impossible to correct the error using any local gate. He argues: "The control of the gate – a physical device – is by modifying some classical variable, which is subject to error. Since one cannot assume infinite precision in any control system, the implications of varying accuracy amongst different gates becomes an important problem. … [I]n certain arrangements a stuck fault cannot be reversed down the circuit stream using a single qubit operator, for it converted a pure state into a mixed state."

Note also that quantum error-correcting techniques cannot correct errors in the probability amplitudes. Consequently, any unknown errors in the implementation of the gates, which are similar in a sense to the unknown and random errors in the variables associated with the control of the gates, are also uncorrectable.

A CNOT based entangled particle generator could be a great help in quantum information experiments. But the elementary two-qubit CNOT gate has not been implemented unconditionally. It is not sufficient to show the correct outputs for the CNOT gate for any specific input values; it should be correct – *and perfectly linear* – for all inputs, but there is not way this can be experimentally established.

This explains why it is notoriously difficult to experimentally demonstrate dense coding or to generate Bell pairs freely [10]. The CNOT circuit

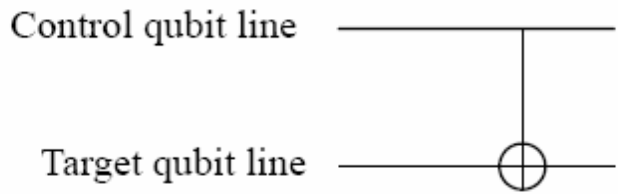

**Figure 1.** A CNOT gate

$$\begin{pmatrix} 1000 \\ 0100 \\ 0001 \\ 0010 \end{pmatrix} \quad (2)$$

would be able to generate correlated states by passing the pure state $\frac{1}{\sqrt{2}}$ *(|0> + |1>)|0>*

$= \frac{1}{\sqrt{2}}$ *(|00> + |10>)* involving two qubits through it, yielding

$$\frac{1}{\sqrt{2}}(|00> + |11>) \quad (3)$$

But the experimental systems available for generation of Bell correlated particles are extremely inefficient [11]. Although there are claims regarding implementation of CNOT gates, we are only sure of the working of the gate for the input probability amplitudes that are 0s and 1s, and there is not enough evidence to establish that the gates work satisfactorily when the input probability amplitudes are other real-valued complex numbers.

## Conclusions

It is argued that there is a basic information theoretic reason, related to the nature of quantum information, which underlies the lack of success in the implementation and control of quantum gates. If quantum information is related exponentially to the number of components in a quantum system, then the realization of the quantum system itself must involve information of exponential complexity. We suggest this is the reason why it has been very difficult to implement unconditionally reliable quantum gates.